**Title:** All Models Are Wrong, But Can They Be Useful? Lessons from COVID-19 Agent-Based Models: A Systematic Review

**Authors and Affiliation:**

Von Hoene, E.[1], Von Hoene, S.[1], Péter, S.A.[1], Hopson, E.[1], Csizmadia, E.E.[1], Fenyk, F.[1], Barner, K.[1], Leslie, T.[1], Kavak, H.[3], Züfle, A.[4] & Roess, A.[2], Anderson, T.[1]

[1]Department of Geography and Geoinformation Science, George Mason University (GMU), 4400 University Drive, Fairfax, Virginia, 22030

[2]Department of Global and Community Health, College of Human and Health Services, George Mason University (GMU), 4400 University Drive, Fairfax, Virginia, 22030

[3]Department of Computational and Data Sciences, George Mason University (GMU), 4400 University Drive, Fairfax, Virginia, 22030

[4]Department of Computer Science, Emory University, 201 Dowman Drive, Atlanta, Georgia, 30322

**Title:** All Models Are Wrong, But Can They Be Useful? Lessons from COVID-19 Agent-Based Models: A Systematic Review

**Abstract**

The COVID-19 pandemic prompted a surge in computational models to simulate disease dynamics and guide interventions. Agent-based models (ABMs) are well-suited to capture population and environmental heterogeneity, but their rapid deployment raised questions about utility for health policy. We systematically reviewed 536 COVID-19 ABM studies published from January 2020 to December 2023, retrieved from Web of Science, PubMed, and Wiley on January 30, 2024. Studies were included if they used ABMs to simulate COVID-19 transmission, where reviews were excluded. Studies were assessed against nine criteria of model usefulness, including transparency and re-use, interdisciplinary collaboration and stakeholder engagement, and evaluation practices. Publications peaked in late 2021 and were concentrated in a few countries. Most models explored behavioral or policy interventions (n = 294, 54.85%) rather than real-time forecasting (n = 9, 1.68%). While most described model assumptions (n = 491, 91.60%), fewer disclosed limitations (n = 349, 65.11%), shared code (n = 219, 40.86%), or built on existing models (n = 195, 36.38%). Standardized reporting protocols (n = 36, 6.72%) and stakeholder engagement were rare (13.62%, n = 73). Only 2.24% (n = 12) described a comprehensive validation framework, though uncertainty was often quantified (n = 407, 75.93%). Over time, reporting of stakeholder engagement and evaluation increased. Studies that claimed policy relevance (n = 354, 66.05%) more often included some evaluation (n = 283, 79.94% vs. n = 125, 68.68%) and stakeholder engagement (n = 61, 17.23% vs. n = 12, 6.59%), though they were less likely to re-use models or share code. Limitations of this review include underrepresentation of non-English studies, subjective data extraction, variability in study quality, and limited generalizability. Overall, COVID-19 ABMs advanced quickly, but lacked transparency, accessibility, and participatory engagement. Stronger standards are needed for ABMs to serve as reliable decision-support tools in future public health crises.

**Keywords:** COVID-19; Agent-based models; Infectious disease modeling; Systematic literature review

**Title:** All Models Are Wrong, But Can They Be Useful? Lessons from COVID-19 Agent-Based Models: A Systematic Review

**Introduction**

The COVID-19 pandemic led to an explosion of computational models ranging from compartmental to agent-based models (ABM) designed to forecast disease trajectories and explore the effect of public health interventions on disease outcomes. However, the push to use models for immediate policy decisions exposed disconnects between modelers' expertise and limited real-world experience as well as user expectations for real-time accuracy and the models' actual capabilities. This frustration was voiced by New York Governor, Andrew Cuomo, who remarked during a press briefing in May 2020: *"Here's my projection model. Here's my projection model. They were all wrong. They were all wrong."* (Cohen, 2020). Such moments reveal how modeling outputs shaped not only research but also high-level decisions and public communication, a perspective echoed in academic critiques. Ioannidis et al. (Ioannidis et al. 2022) argue bluntly that COVID-19 forecasting "failed", citing the large discrepancies between model projections and observed outcomes while Jewell et al. (Jewell et al. 2020) caution that overly confident forecasts misled decision-makers when uncertainty was not adequately communicated. Jewell et al. (Jewell et al. 2020) write *"This appearance of certainty is seductive when the world is desperate to know what lies ahead."* The lack of alignment between COVID-19 projection models with observed outcomes undermined public confidence and sparked a debate about the appropriate use of simulation models in public health decision-making (Holmdahl & Buckee 2020; Ioannidis et al. 2022; Squazzoni et al. 2020).

While models can support timely interventions, flawed or misunderstood models may appear alarmist, undermine trust, and lead to overcorrections that do more harm than good. For example, the highly influential Institute for Health Metrics and Evaluation (IHME) COVID-19 model was used at both federal and state levels to justify policy interventions that shut down parts of the US economy (Robbins 2021). However, the model assumed uniform compliance with social distancing policies, producing inaccurate estimates with narrow uncertainty bands (Jewell et al. 2020; Schroeder 2021). IHME specifically has long been criticized for lack of transparency in their methods for generating global health metrics (Shiffman & Shawar 2020). Kitching et al. (Kitching et al. 2006) remind us that this tension is not unique to COVID-19 simulations. A widely cited public health example is the case of the 2001 UK foot-and-mouth disease outbreak, in which a model assumed transmission was possible between adjacent farms (Ferguson et al. 2001). This assumption led the model to forecast a much larger and faster-growing outbreak than occurred, prompting aggressive culling of millions of livestock. The response caused severe economic losses, long-term environmental impacts, and eroded public trust in science-driven policy (Anderson 2021; Kitching et al. 2006).

Drawing on classic ideas from statistician George Box (Box 1979), Holmdahl and Buckee (Holmdahl & Buckee 2020) reiterate that despite being "wrong", models can still be useful. While no model can capture the full complexity of the real world, they have the potential to offer valuable insights from estimating important disease parameters (e.g. $R_0$, $R_E$) to providing guidance on interventions. For example, government organizations used models to forecast the size of the 2014 Ebola outbreak and the geographic patterns of spread in West Africa, allowing decision-makers to prioritize where to direct resources and improve surveillance (Fischer 2016; Martin I. Meltzer, PhD1 et al. 2014; WHO Ebola Response Team 2014). In another example, models were used to estimate the $R_0$ heterogeneity of a 2008 outbreak of cholera in Zimbabwe, shaping vaccine and clean-water deployment plans (Mukandavire et al. 2011). Despite uncertainty in COVID-19 model forecasts, the *COVID-19 Forecast Hub* (Cramer et al. 2022), provided ensemble forecasts that combined multiple models, resulting in improved accuracy with clearer communication of uncertainty, and informed decisions by organizations like the CDC in their official COVID-19 guidance.

The COVID-19 pandemic has reignited long-standing questions about what makes models *useful* for guiding public health policy decisions. Scholars argue that models fail not only when they are inaccurate, but when their assumptions, scope, and intended uses are unclear (Anderson 2021; Ioannidis et al. 2022; Saltelli et al. 2020). Multiple recent commentaries and major calls for reform in the modeling community have highlighted common principles of making disease simulation models truly useful. Saltelli et al. (Saltelli et al. 2020) offer a manifesto: models should be transparent about assumptions and limitations, openly shared, and co-produced with stakeholders; they should clarify uncertainty and avoid false precision. Squazzoni et al. (Squazzoni et al. 2020) echo these ideas with a call for open-source software and tools, adoption of standard protocols for model documentation, and the use of permanent online repositories of model code to speed up assessment and model re-use. Eker (Eker 2020) stresses that although COVID-19 models can be useful in many ways, they require validation and a clear communication of their uncertainties to be credible for decision-making. Squazzoni (Squazzoni et al. 2020) warns that a model that is not rigorously checked could have serious consequences in the findings are used to inform decision making.

The COVID-19 pandemic offered a rare opportunity to reflect on how simulation models support public health decision-making, revealing both advances in modeling practice and persistent gaps that limit their utility. Among the different simulation approaches, agent-based models (ABMs) are broadly described as being a useful tool for decision-support (Bui & Lee 1999; DeAngelis & Diaz 2019; Klabunde & Willekens 2016; Sengupta & Bennett 2003). Epidemiological ABMs offer unique insights due to their ability to simulate complex interactions and behaviors between heterogeneous individuals (*i.e. agents*) and their environments from which disease outcomes emerge and are well-suited for exploring "what if" scenarios, such as alternative policy interventions (e.g., 24–26). In this paper, we conducted a systematic review of COVID-19 ABM studies published between 2020 and 2023 to answer the following research question: "*To what extent did COVID-19 ABMs adhere to best practices of transparency and re-use, interdisciplinary collaboration and stakeholder engagement, and model evaluation — and how did these practices evolve over time?*"

Drawing on expert critiques from the literature (see references in Table 1), we assess each study against a set of nine criteria widely considered essential for making disease simulation models "useful": (i) use of open source ABM tools for model development, (ii) adoption of standard protocols for model communication, (iii) clear acknowledgment of model assumptions and limitations, (iv) use of permanent model repositories for model code, (v) building upon existing models, (vi) collaboration across multiple disciplines, (vii) inclusion of stakeholders in the modeling cycle, (viii) implementation of model evaluation, and (ix) quantification of model uncertainty.

The remainder of this paper is structured as follows. We first describe the methodology used to search, retrieve, screen, and assess relevant studies. We then present the results of our systematic review and discuss key trends, gaps, and implications. Finally, we conclude by outlining limitations and directions for future research.

**Methodology**

To address our research question, we conducted a systematic review of published COVID-19 ABM studies, following established Preferred Reporting Items for Systematic reviews and Meta-Analyses (PRISMA) guidelines for transparency and reproducibility (Page et al. 2021) (see Supplement 5 and 6). The methodology consisted of three stages: (i) literature search and retrieval, (ii) screening and eligibility assessment, and (iii) data extraction and analysis.

*Assessment Framework and Criteria Definitions.* Nine criteria, falling more broadly under the categories of "Model Transparency and Re-Use", "Interdisciplinary Collaboration and Stakeholder Engagement", and "Model Evaluation", were derived from expert commentaries published in 2020-21 that critiqued the application of COVID-19 models in practice. A criterion was included in our framework if discussed in at least two separate commentaries. Table 1 lists and defines the final set of criteria used in our assessment framework, along with a brief rationale for their inclusion.

*Table 1. Assessment framework and criteria definitions.*

| Category | Criteria | Description and Rationale |
|---|---|---|
| Model Transparency and Re-Use | Use of open-source ABM tools for model development | Models implemented in publicly available agent-based modeling platforms (e.g., NetLogo, GAMA, Julia, Repast4Py, Mason) promote access for reviewers and replication (Hunter & Kelleher 2021; Squazzoni et al. 2020). |
| | Adoption of standard protocols for model communication | Use of standardized reporting frameworks (e.g., ODD (Grimm et al. 2010), ODD+D (Müller et al. 2013)) ensures clear and consistent documentation, making models easier to understand, compare, and replicate (Grimm et al. 2020; Squazzoni et al. 2020). |
| | Clear acknowledgment of model assumptions and limitations | Explicitly stating assumptions and limitations defines the scope of a model and increases transparency (Eker 2020; Saltelli et al. 2020). |
| | Model downloadable from permanent repository | Making model code available in permanent repositories (e.g., GitHub, Zenodo) supports re-use, assessment, and long-term accessibility (Ioannidis et al. 2022; Squazzoni et al. 2020). |
| | Re-use or extension upon existing models | Extending or reusing prior models reduces redundancy, fosters comparability across studies, and enhances credibility by building on validated frameworks (Grimm et al. 2020; Squazzoni et al. 2020). |
| Interdisciplinary collaboration and stakeholder engagement | Interdisciplinary collaboration | Including expertise across multiple disciplines ensures models capture biological, technical, and social dimensions of disease spread (Ioannidis et al. 2022; Squazzoni et al. 2020). |
| | Inclusion of stakeholders in the modeling cycle | Involving policymakers, public health practitioners, or community representatives ensures model relevance and increases the likelihood of model uptake (Saltelli et al. 2020; Squazzoni et al. 2020). |
| Model evaluation | Implementation of model evaluation | Verification, calibration, sensitivity analysis, and validation assess credibility and robustness, increasing trust in the model as a decision-support tool (Eker 2020; Squazzoni et al. 2020). |
| | Model uncertainty quantification | Explicitly measuring and communicating uncertainty (e.g., through confidence intervals, sensitivity testing) |

|  |  | acknowledges model limits and avoids false precision (Ioannidis et al. 2022; Saltelli et al. 2020). |
|---|---|---|

***Search strategy and retrieval***: Based on Zhang et al. (2023), we considered the following academic web portals and databases for our search, including: PubMed, MEDLINE, EMBASE, Web of Science, Scopus, Science-Direct, EBSCO, Wiley, and WHO COVID-19 Database. MEDLINE was excluded since it draws from PubMed. Science-Direct was excluded since it limits the number of Boolean operators in a search. WHO COVID-19 Database was excluded since it stopped operation in June 2023. EMBASE and Scopus were excluded due to lack of organizational access.

Due to these exclusions, articles were retrieved from Web of Science, EBSCO, PubMed, and Wiley on 1/30/2024 using a four-year date range of 01/01/2020-12/31/2023 to capture all COVID-19 related models. Following Zhang et al. (2023), the search query required one or more of the terms 'agent-based models', 'agent-based modelling', 'individual-based model', 'multi-agent system', 'coronavirus disease 2019', 'COVID-19', 'COVID-2019', 'severe acute respiratory syndrome coronavirus 2', 'SARS-CoV2' and 'SARS-CoV-2'. The search strategy was intentionally designed to include specific disease-related terms (e.g., 'coronavirus disease 2019') rather than broader terms (e.g., 'disease outbreaks'), allowing us to efficiently retrieve articles most directly relevant to our criteria. This specificity minimized the retrieval of irrelevant or overly general records, thereby increasing the precision of our literature search. A total of 902 articles were retrieved, as follows: Web of Science (n = 893), PubMed (n = 3), Wiley (n = 6), EBSCO (n = 0). In general, PubMed, Wiley and EBSCO failed to return articles at the intersection of the ABM approach and COVID-19. For example, PUBMED yielded 17,489 articles using keywords related to COVID-19 and 171 articles using keywords related to ABMs, the intersection of these two searchers returned only 3 articles. We include the search history for the databases in Supplement 1. All supplementary material for this study can be found at https://tinyurl.com/COVID-ABM-Review-Supplement. Of the 902 articles, 32 were excluded since they were not conference proceedings or journal articles and 10 were marked as duplicates by automation tools, resulting in 860 records for title and abstract screening (Figure 1).

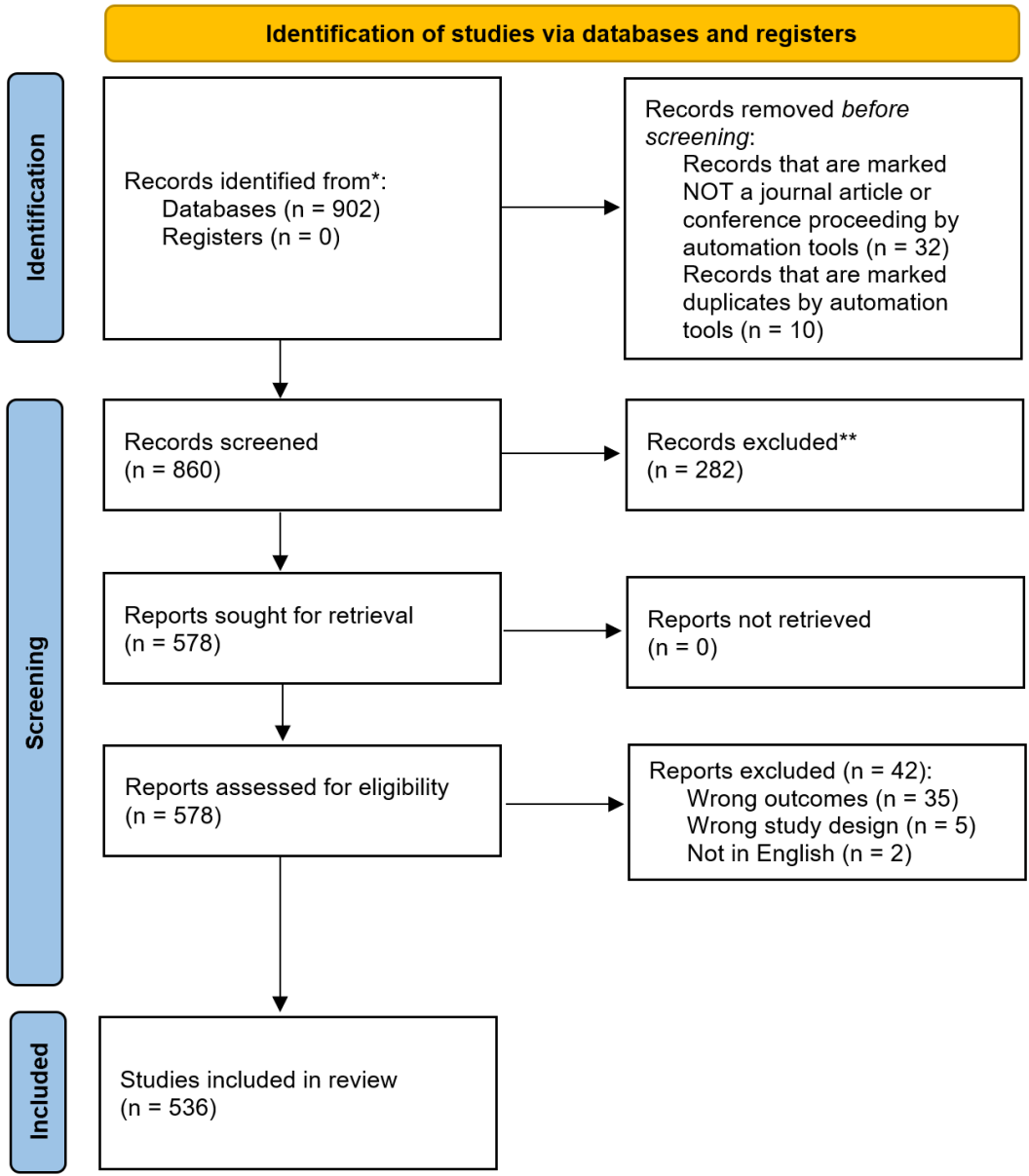

*Figure 1. PRISMA 2020 flow diagram outlining our systematic review protocol (adapted from Page et al. 2021).*

***Title and Abstract Screening.*** Rayyan systematic review software (rayyan.ai) was used to manage the initial screening of the 860 articles, which relied on article titles and abstracts to tag articles for inclusion or exclusion. At each stage, two authors (E.V.H. and T.A) screened each article independently, using the inclusion and exclusion criteria as described in Table 2. There were 97 conflicts (87% agreement), which were resolved through discussion. Through this title and abstract screening, we excluded 282 articles and included 578 articles for full text assessment (Figure 1).

***Full Text Assessment.*** Covidence systematic review software (app.covidence.org) was used for the full text assessment for eligibility to be included in the review. At least two authors (T.A. and several rotating second reviewers including F.F., S.P., K.B., E.C., and E.H.) assessed the full text of each article independently. There were 7 conflicts (98% agreement), which were resolved by author E.V.H. Through the full text assessment and the inclusion and exclusion criteria described in Table 2, we excluded 42 articles and included a final 536 articles for data extraction (Figure 1). The list of the 536 articles can be found in Supplement 2.

*Data Extraction.* Each article was assigned to two independent reviewers for data extraction (F.F., S.P., K.B., E.C., S.V.H., and E.H.). The data extraction form can be found in Supplement 3. Across all pairs of reviewers, there was an average agreement of 74% (min: 68.8%, max: 83.8%). Conflicts were resolved by either T.A. or E.V.H.

*Table 2. The inclusion and exclusion criteria.*

| Reason | Include | Exclude |
|---|---|---|
| *Outcome* | The study presents or describes a model that simulates the transmission of COVID-19 from one individual to another. Studies that model the transmission of disease often utilize parameters to model the spread of the disease and should be included (disease staging including susceptible, infectious, recovered, length of time in each stage, transmission likelihood etc.). | The model simulates the transmission of other diseases, outside of COVID-19. Studies that measure risk as a proxy for likelihood of COVID-19 transmission should be excluded. This is typically the case in very small-scale simulations (e.g. simulating the boarding of a plane). |
| *Study Design* | The modeling approach is an agent-based model. Approaches with other names equivalent to an "agent-based model" such as "multi-agent", "individual-based model" should be included. If the smallest unit is an individual, it is likely an agent-based model. Hybrid models where one component of the model is an agent-based model should be included. | Other common models that are solely machine learning models or compartmental models should be excluded. |
| *Other Consideration* | The article is either a conference article or a journal article. The article should be written in English. | Review articles should be excluded. Articles not in English should be excluded. |

**Results and Discussion**

*Overview.* Our systematic review identified and extracted data from 536 COVID-19 ABM studies published between January 2020 and December 2023. The full dataset is openly accessible in Supplement 7 at https://tinyurl.com/COVID-ABM-Review-Supplement. The first COVID-19 ABM study in our collection was published on May 20, 2020. Figure 2 shows the quarterly publication trends of COVID-19 ABM studies from 2020 through 2023. Publications peaked in July – September 2021, reflecting a lag between early pandemic model demand and their appearance in published literature. Although models are needed most urgently at the outset of an epidemic's outset, the time required for development, calibration, peer review and publication processes may delay availability until after the critical decision-making window has passed.

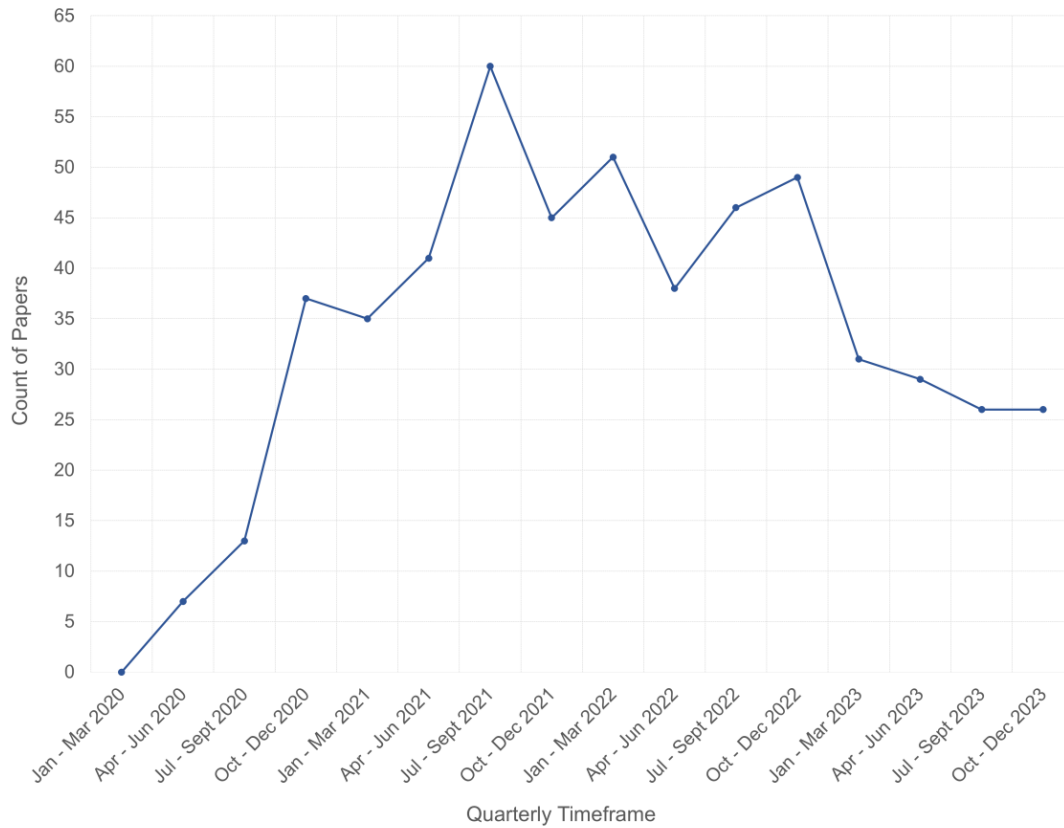

*Figure 2. Count of COVID-19 ABM studies per quarter from 2020 to 2023.*

      The COVID-19 ABM studies were published across 261 publication outlets including 219 journals, 2 books, and 40 conference proceedings. The top three most common publication outlets were *Nature Scientific Reports* (36 studies), *PLoS ONE* (31 studies), *PLoS Computational Biology* (16 studies). Among the 219 journals, 103 were fully open access (OA), 112 followed a hybrid model (subscription with OA options), and 4 were subscription only. We note that of the 103 OA journals, 88 were listed in the Directory of Open Access Journals (DOAJ), regarded as a "whitelist" of OA journals, and 15 journals (totaling 30 studies) were not. This may suggest variability in the quality of the studies included in our analysis.

      Geographically, the countries where COVID-19 ABM case studies were undertaken spanned all continents excluding the poles. The largest concentrations were models of the United States (n = 109, 20.34%), United Kingdom (n = 50, 9.33%), China (n = 49, 9.14%), Australia (n = 27, 5.04%), Canada (n = 25, 4.66%), Italy (n = 25, 4.66%), France (n = 18, 3.36%), and Germany (n = 18, 3.36%), which together accounted for a disproportionately high share of case studies (Figure 3).

      Across the 536 studies, the predominant purpose of the COVID-19 ABMs (Figure 4) was to understand the role of specific factors or interventions in disease transmission and control (n = 383, 71.46%). A majority of studies assessed the effects of policy guidelines and health behaviors (n = 294, 54.85%; e.g., mask mandates, school closures, or vaccine uptake), followed by contact networks (n = 26, 4.85%; e.g., spread across different network types, such as workplaces, schools, long-term care facilities, families, etc.), spatial effects (n = 18, 3.36%; e.g. urban vs. rural transmission patterns), disease

parameters (n = 16, 2.99%; e.g., variant-specific dynamics), mobility (n = 11, 2.05%; e.g., influence of varying travel patterns), population structure (n = 11, 2.05%; e.g., introducing heterogeneity into agent populations), and the spread of misinformation or information (n = 7, 1.31%; e.g., how information impacts health behavior uptake). A smaller share of studies (n = 63, 11.75%) proposed new modeling frameworks to simulate the spread of COVID-19 in specific locations or settings (n = 32, 5.97%; e.g., hospitals, university campuses), to combine ABM with other simulation approaches (n = 16, 2.99%; e.g. hybrid ABM-discrete event simulation), and to generically model disease spread (n = 15, 2.80%; e.g., FRED (Grefenstette et al. 2013), Covasim (Kerr et al. 2021)). Another 15.11% (n = 81) developed or refined methods for the following: advancing disease modeling (n = 57, 10.63%; e.g., a framework for implementing health behaviors), assessing model outcomes (n = 17, 3.17%; e.g., strategies for calibrating, sensitivity analysis and validation), and improving computational efficiency (n = 7, 1.31%; e.g., optimizing ABM to simulate faster with more agents). Only 1.68% (n = 9) of studies aimed to produce epidemic forecasts under existing conditions, reflecting the challenges of real-time calibration in rapidly evolving conditions with little data availability. Overall, the distribution of study purposes suggests that ABMs during COVID-19 were developed as exploratory tools to test the impact of interventions, examine behavioral and structural drivers of disease spread, and innovate on ABM methodology.

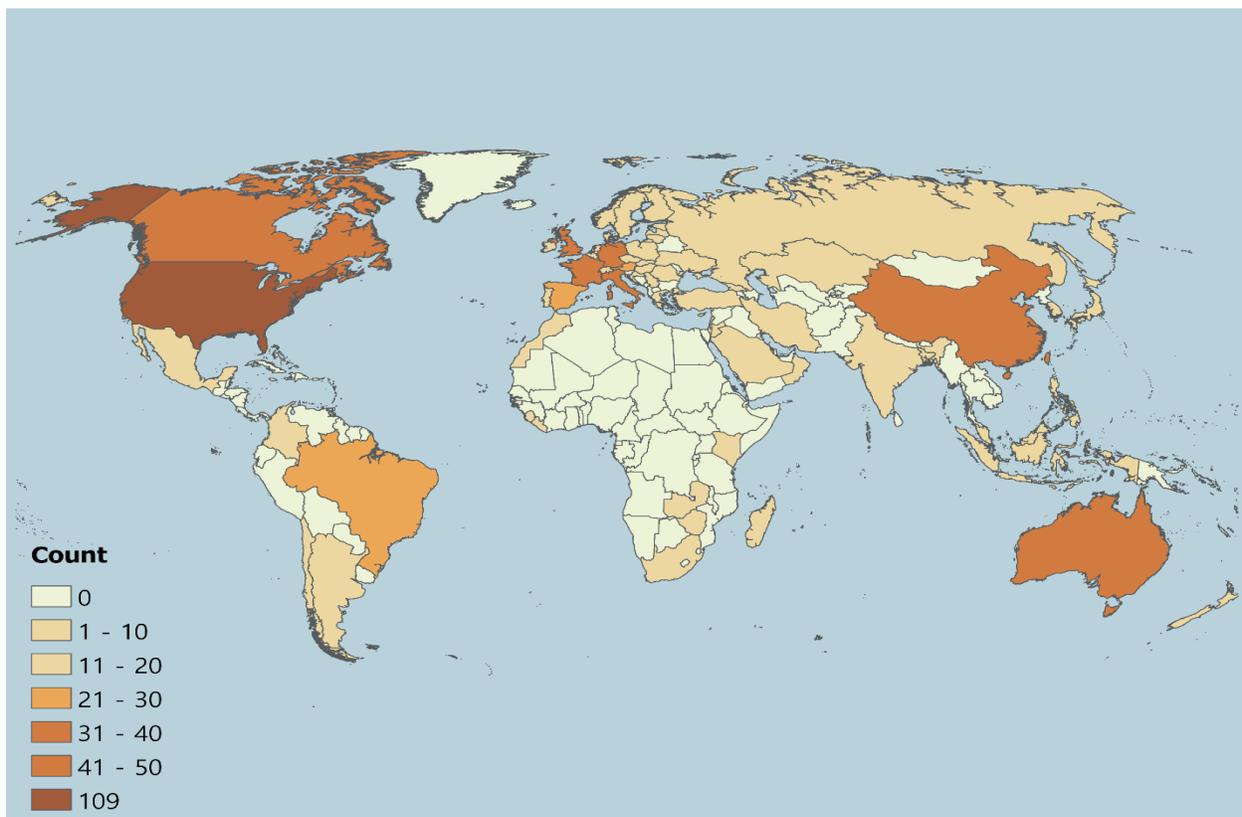

*Figure 3. Geographic distribution of COVID-19 ABMs by study country. Counts represent the number of studies in which the model was applied to simulate disease spread in that country. The map was created in ArcGIS Pro, and the generalized world boundary was obtained from ESRI's open-source ArcGIS Online Portal ([https://www.arcgis.com/home/item.html?id=2b93b06dc0dc4e809d3c8db5cb96ba69](https://www.arcgis.com/home/item.html?id=2b93b06dc0dc4e809d3c8db5cb96ba69)).*

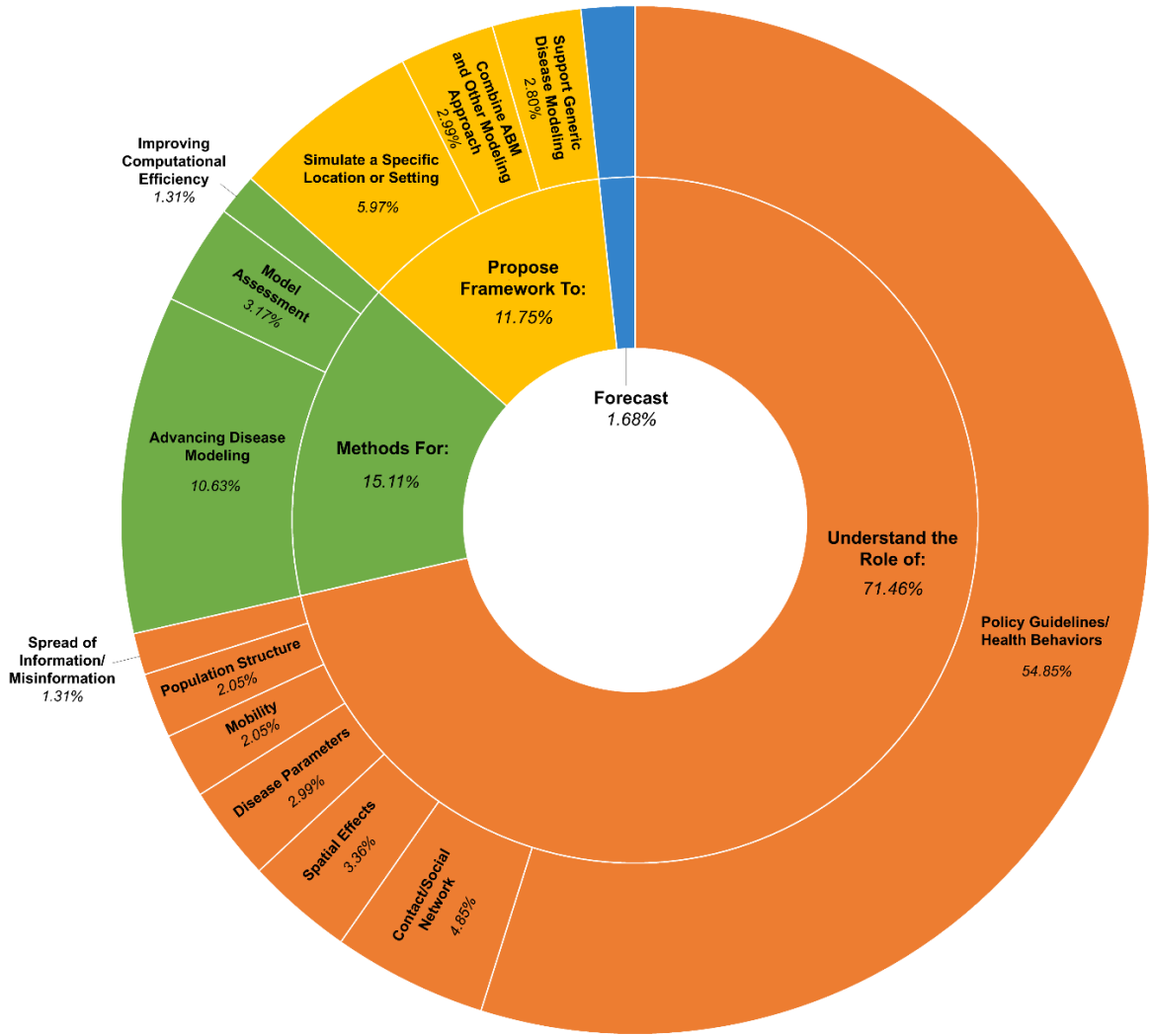

*Figure 4. Percentages of COVID-19 ABM studies by purpose (inner pie chart) and by sub-purpose (outer pie chart).*

Two-thirds of the studies (n = 354, 66.05%) explicitly stated that their models or model outputs are "useful" for informing public health policy making or other researchers, highlighting the field's emphasis on practical application. The implication that COVID-19 ABMs will be used for decision making support places a responsibility on the modeling community to use best practices in model development, documentation, and evaluation. Table 3 provides an overview of the assessment of the COVID-19 ABM studies against a set of nine criteria that are considered essential for translating models into useful tools for decision-making, falling more broadly under the categories of "Model Transparency and Re-Use", "Interdisciplinary Collaboration and Stakeholder Engagement", and "Model Evaluation". Analyses are presented for all studies, for studies claiming potential policy usefulness versus non-claiming studies (policy-claiming, n = 354, 66.05%; non-claiming, n = 182, 33.95%), and by publication year from 2020 to 2023 (2020: n = 58, 10.82%; 2021: n = 181, 33.77%; 2022: n = 184, 34.33%; 2023: n = 113, 21.08%). Percentages reported in Table 3 are calculated relative to the number of papers in each subset: for all studies, percentages are out of 536; for policy-claiming versus non-claiming studies, percentages are out of 354 and 182, respectively; and for each publication year, percentages are out of the total papers published that year. Below, we present and discuss these results. Supplement 4 provides a more detailed sub-analysis of the criteria.

*Table 3. Assessment of COVID-19 ABM studies against nine criteria essential for decision-support modeling. Criteria are analyzed across all articles, articles claiming potential usefulness, and by publication year (2020–2023). Percentages are calculated relative to the total papers in each subset (see text for details).*

| Criteria (%) | Overall<br>N = 536 | Policy Claiming<br>N = 354 | Non-Claiming<br>N = 182 | 2020<br>N = 58 | 2021<br>N = 181 | 2022<br>N = 184 | 2023<br>N = 113 |
|---|---|---|---|---|---|---|---|
| **MODEL TRANSPARENCY AND RE-USE** | | | | | | | |
| **Use of Open-Source ABM Tools for Model Development** | 19.78 | 22.32 | 14.84 | 25.86 | 18.23 | 20.11 | 18.58 |
| **Adoption of Standard Protocol for Model Communication** | 6.72 | 8.19 | 3.85 | 5.17 | 4.97 | 8.15 | 7.96 |
| **Clear Acknowledgement of Model Assumptions and Limitations** | 62.31 | 65.25 | 56.59 | 53.45 | 63.54 | 59.24 | 69.91 |
| Clear model assumptions | 91.6 | 91.53 | 91.76 | 87.93 | 91.16 | 92.93 | 92.04 |
| Clear model limitations | 65.11 | 67.80 | 59.89 | 53.45 | 66.30 | 62.50 | 73.45 |
| **Model Downloadable from Permanent Repository** | 40.86 | 38.70 | 45.05 | 39.66 | 45.86 | 36.41 | 40.71 |
| **Re-use or Extension Upon Existing Models** | 36.38 | 33.62 | 41.76 | 24.14 | 33.70 | 40.76 | 39.82 |
| **INTERDISCIPLINARY COLLABORATION AND STAKEHOLDER ENGAGEMENT** | | | | | | | |
| **Inclusion of Stakeholders in the Modeling Cycle** | 13.62 | 17.23 | 6.59 | 5.17 | 13.26 | 14.67 | 16.81 |
| **Interdisciplinary Collaboration: More than one discipline per study** | 63.25 | 64.41 | 60.99 | 51.72 | 62.98 | 66.85 | 63.72 |
| **MODEL EVALUATION** | | | | | | | |
| **Implementation of Model Evaluation: Verification or Calibration or Sensitivity Analysis or Validation** | 76.12 | 79.94 | 68.68 | 58.62 | 75.69 | 79.35 | 80.53 |
| Describe verification | 13.06 | 14.69 | 9.89 | 6.90 | 11.05 | 15.76 | 15.04 |
| Describe calibration | 49.07 | 52.82 | 41.76 | 39.66 | 49.17 | 54.35 | 45.13 |
| Describe sensitivity analysis | 38.99 | 41.53 | 34.07 | 29.31 | 40.88 | 38.59 | 41.59 |
| Describe validation | 35.26 | 40.96 | 24.18 | 22.41 | 35.91 | 33.70 | 43.36 |
| **Model Uncertainty Quantification** | 75.93 | 79.10 | 69.78 | 68.97 | 73.48 | 82.07 | 73.45 |

***Model Transparency and Re-Use.*** Overall, only 19.78% (n = 106) of models were implemented with open-source ABM software, with NetLogo being the most common platform (n = 57, 10.63% of total studies), followed by GAMA (n= 13, 2.43%), and Julia (n = 10, 1.87%). In contrast, most studies relied on custom-built code in general-purpose programming languages such as Python, C++, and Java. As Squazzoni et al. (Squazzoni et al. 2020) note, reliance on open-source platforms and tools minimizes barriers to replication and re-use, which may help explain why only 36.38% (n = 195) of studies explicitly built upon existing models. One exception is Covasim (Kerr et al. 2021), developed in Python and openly shared with extensive documentation, which emerged as one of the most frequently reused frameworks (n = 30, 5.60%,). This suggests that re-use depends not only on whether a model is implemented in open-source ABM software, but also on accessibility and transparency.

      On that note, adoption of standard protocols for model documentation was rare (n = 36, 6.72%), with the Overview Design and Details (ODD) protocol (Grimm et al. 2010) accounting for most (n = 31, 5.78%). More detailed frameworks such as ODD+D (Decisions) (Müller et al. 2013) and International Society for Pharmacoeconomics and Outcomes Research Society for Medical Decision Making (ISPOR-SMDM) Modeling Good Research Practices (Caro et al. 2012) were almost entirely absent (<1%). Clear articulation of model assumptions was common (n = 491, 91.60%), yet explicit statements of limitations were provided in only 65.11% (n = 349) of studies. Although 62.31% (n = 334) of studies included a description of both assumptions and limitations, indicating that a substantial fraction of the models was presented without fully disclosing model scope. Less than half (n = 219, 40.86%) made their model code available for download, most frequently via GitHub (n = 169, 31.53%), followed by Zenodo (n = 18, 3.36%) and Gitlab (n = 10, 1.87%), limiting transparency, assessment, and further development of model code. To promote transparency, facilitate reproducibility, and encourage model reuse, we provide a compiled list of available model code links from the 536 reviewed articles in Supplement 8 at https://tinyurl.com/COVID-ABM-Review-Supplement. Given repeated calls in the literature for increased model transparency, the low uptake of open-source ABM software, standard protocols for model documentation, and availability of model code may reflect systematic issues related to time pressure, the cost burden (e.g. money, time, effort), or lack of awareness. Furthermore, while several journals recommend using such tools, it is not typically a requirement.

***Interdisciplinary Collaboration and Stakeholder Engagement.*** Most studies (n = 339, 63.25%) included authors from more than one discipline, with 28.92% (n =155) involving 3 or more (Supplement 4), which reflects some alignment with best-practice recommendations for interdisciplinary collaboration. Disciplines were classified according to the U.S. National Science Foundation (NSF) Codes for Classification for Research (University of New Mexico, Office of Sponsored Programs, 2025). We also added a category for industry/non-academic fields (e.g., government organizations or contractors, hospitals, churches). The x-axis of Figure 5 shows each discipline alongside the number and percentage of papers (out of 536) that included at least one author from that discipline. The most represented disciplines included life sciences (n = 252, 47.01%), industry/non-academic fields (n = 208, 38.81%), computer and information sciences (n = 167, 31.16%), and engineering (n = 137, 25.56%). The interdisciplinary collaborations between discipline pairs (i.e., number of papers with at least one author from the two disciplines) are shown in each cell in Figure 5. For each pair, a chi-square test ($\chi^2$) compared the observed number of papers (the first bolded value in each cell) coauthored by the two disciplines to the expected number ($\hat{n}$) under random collaboration. Cells colored in gray indicate that there is no more or less collaboration than there would be at random (p-value greater than 0.05). Colored cells (also denoted with an * after each cell's parentheses) indicate statistically significant deviations: orange represents more collaborations than expected, and blue indicates fewer.

Collaborations that occurred more frequently than expected and were statistically significant included life sciences (which include biomedical sciences and health sciences) and industry/non-academic fields (n = 120, $\hat{n}$=97.8, $\chi^2$ = 15.6) followed by life sciences and mathematics and statistics (n = 67, $\hat{n}$=52.7, $\chi^2$ = 9.3). Notably, while authors from engineering appeared in about 25% of studies (more than mathematics and statistics), collaborations with life sciences occurred less frequently than expected (n = 46, $\hat{n}$=64.4, $\chi^2$ = 13.3), although possibly engineering has less of a stake in disease simulations. It is important to note that the chi-square analysis relies on an implicit assumption: the expected number of collaborations is calculated as if each paper had exactly two authors, with each author randomly paired with one co-author. In practice, papers may have only one author or many co-authors, which can increase or decrease the observed number of collaborations relative to this expectation. As a result, some observed patterns, particularly in fields with many multi-author papers (e.g., life sciences), may reflect differences in author count rather than true disciplinary preference.

Only 13.62% (n = 73) of studies reported direct stakeholder involvement in the modeling cycle. Of these studies, the most common stakeholder roles were to inform model assumptions (n = 23, 4.29%), guide model intervention scenarios (n = 21, 3.92%), use model results for decision making (n = 16, 2.99%), and provide expert validation (n = 12, 2.24%). Less frequent roles included model testing and calibration (n = 3, 0.56%), providing data (n = 3, 0.56%), supporting dissemination of model results (n = 2, 0.37%), further analyzing the model data (n= 2, 0.37%), and defining performance metrics (n = 1, 0.19%). Stakeholders ranged from institutional actors including city, county, and state governments (e.g., "City of Chicago, Cook County, and State of Illinois"), public health agencies (e.g. "National Health Service in the United Kingdom", "Health Canada", "Polish Ministry of Health"), schools or university administration (e.g. "UC San Diego Return to Learn Program") to individual actors including public health officials, educators, healthcare providers, business owners, local planners, and leaders of worship.

Yet, in general, stakeholder participation is strikingly low given that two-thirds of studies claimed potential policy relevance. Yet, this aligns with literature that points to a longstanding gap between modelers and policy makers (Mihaljevic et al. 2024). The lack of engagement can be attributed to several practical and political challenges: models can seem like "black boxes" that are hard to trust, policies often change quickly and unpredictably, decisions must be made on shorter timelines than models can support, government contracting processes can be slow and inflexible, political values can outweigh evidence, and there are usually many different stakeholders to involve (Gilbert et al. 2018). As Edmonds et al. (Edmonds et al. 2019) states: a model can only be validated relative to a clearly defined purpose, and if stakeholders are not involved in setting that purpose, it is difficult to claim validity for policy or operational decision-making.

*Figure 5. Matrix of interdisciplinary collaboration across the 536 papers assessed in this study. Within each cell, bold values indicate the observed number of papers with at least one author from both disciplines. ñ denotes the expected number of papers under random collaboration, while $\chi^2$ represents the corresponding chi-square statistic. Colored cells and an asterisk (\*) denote statistically significant differences ($p < 0.05$): blue indicates fewer collaborations than expected, orange indicates more collaborations than expected, and grey indicates non-significant results. The x-axis includes the count and percentages of papers with at least one author from the given discipline.*

***Model Evaluation.*** Overall, 76.12% (n = 408) of studies mention at least one evaluation method: verification, calibration, sensitivity analysis, or validation. Calibration was the most common (n = 263, 49.07%), followed by sensitivity analysis (n = 209, 38.99%), and validation (n = 189, 35.26). Model verification was reported less frequently (n = 70, 13.06%), which may reflect its integration into routine model development and is not explicitly reported as part of the model evaluation. While most COVID-19 ABM studies incorporated some form of evaluation, only 2.24% (n = 12) described a systematic framework encompassing verification, calibration, sensitivity analysis, and validation together. Communication of either aleatory or epistemic model uncertainty was more frequent (n = 407, 75.93%), typically through repeated runs with confidence intervals or sensitivity analysis, respectively. Taken together, these findings highlight a tension between the urgency of producing models and the time-intensive process of rigorous evaluation. Given that two-thirds of studies positioned themselves as potentially useful for public health policy, the uneven application of evaluation methods raises concerns about the credibility and reproducibility of their insights. Strengthening evaluation practices remains essential if ABMs are to serve as reliable decision-support tools in future pandemics.

***Trends Over Time.*** Table 3 shows the patterns of adherence to best practices in COVID-19 ABMs over time (2020-2023). Note that percentages are calculated relative to the total number of studies published each year. We expected that because modeling standards evolved rapidly during the early stages of the pandemic, earlier studies may be less likely to meet the criteria. Transparency indicators remained low overall: the use of open-source ABM tools fluctuated between 18-25% without a clear upward trajectory, while the adoption of standard communication protocols were consistently adopted by less than 9% of studies each year. Reporting of model assumptions or limitations strengthened over time. Assumptions remained consistently well documented (>87%), but explicitly describing the model limitations rose from 53.45% (n = 31) in 2020 to over 73.45% (n = 83) in 2023. Availability of models for download hovered around 36-46% without consistent gains. Evidence of re-use improved modestly whereby studies building on existing models increased from about 24% in 2020 to nearly 40% by 2022-23.

Interdisciplinary collaboration remained steady (about 63-67%) between 2021-2023 after a weaker start in 2020 (51.72%). Stakeholder engagement, while limited overall, grew from 5% in 2020 to almost 17% in 2023, suggesting gradual recognition of the value of participatory modeling. Evaluation practices showed the clearest progress. The share of studies applying verification, calibration, sensitivity analysis, or validation rose approximately from 59% in 2020 to 81% in 2023. Calibration generally remained consistent over time, but increased 10% between 2020 and 2021. Validation more than doubled from 22.41% (n=13) to 2020 to 43.36% (n = 49) in 2023. Communication of uncertainty was consistently strong (69-82%), peaking in 2022. In general, we find that while COVID-19 ABM development strengthened their position in model evaluation and stakeholder engagement over time, gaps in transparency, documentation, and open accessibility persist.

***Policy-Claiming vs. Non-Claiming Studies.*** Of the 536 COVID-19 ABM studies, 354 (66.05%) explicitly claimed policy relevance, while 182 (33.95%) did not. Percentages reported below are calculated relative to the number of studies in each group. Comparing these groups shows that, although policy-claiming studies were somewhat more likely to meet best-practice criteria, the differences were modest and inconsistent. For example, policy-claiming studies were slightly more likely to use open-source ABM tools (n = 79/354, 22.32% vs. n = 27/182, 14.84%), follow standard communication protocols (n = 29 8.19% vs. n = 7 3.85%), and acknowledge assumptions and limitations (n = 231, 65.25% vs. n = 103, 56.59%), although overall adoption of existing models remained low. Surprisingly, non-claiming studies were more likely to make code available (n = 82, 45.05% vs. n = 137, 38.70%) and to build upon existing models (n = 76, 41.76% vs. n = 119, 33.62%). Interdisciplinary authorship was common in both groups (n = 228, 64.41% vs. n = 111, 60.99%). Stakeholder involvement, however, was more than twice as frequent among policy-claiming studies (n = 61, 17.23% vs. n = 12, 6.59%). The clearest differences emerged in evaluation. Policy-claiming studies more often reported verification, calibration, sensitivity analysis, or validation (n = 283, 79.94% vs. n = 125, 68.68%), particularly validation (n = 145, 40.96% vs. n = 44, 24.18%), which is critical for model credibility. These findings suggest that while many policy-oriented studies took additional steps to strengthen credibility, there remain opportunities to better align claims of usefulness with best-practice standards.

**Limitations and Future Work**

This review has several limitations that should be acknowledged. First, by focusing exclusively on published journal and conference articles, we may have excluded relevant models described in preprints, reports, or those that remain unpublished. Second, the focus on English-language publications may have led to an overrepresentation of studies from English-speaking countries (Figure 3). Third, we examined only ABMs, meaning that findings cannot be generalized to the broader COVID-19 modeling community.

Future work could extend this review to include other modeling approaches (e.g., compartmental, machine learning) to determine whether the observed gaps are unique to ABMs or common across simulation methodologies. Fourth, 30 studies included in our analysis were published in journals that were both OA and not listed in the DOAJ, suggesting variability in study quality. Future work may further investigate whether OA journals participate in activities considered "predatory" and conduct a sub-analysis of these studies. Finally, while our nine criteria were grounded in expert commentary, the data extraction process inevitably involved some subjectivity, with variability across reviewers. Despite these limitations, this review offers a longitudinal analysis on ABM modeling practices during the COVID-19 pandemic, highlighting both important progress and persistent gaps. Future research could build on these insights by identifying barriers that undermine best practices and by continued development and promotion of frameworks, tools, and participatory modeling approaches that strengthen them.

**Conclusion**

This review assessed the extent to which COVID-19 ABMs published between 2020 and 2023 adhere to best practices of transparency and re-use, interdisciplinary collaboration and stakeholder engagement, and model evaluation. Among more than 500 studies, models were widely used to explore the effects of interventions with some studies reporting that their models were used directly in practice. However, in general, models fell short of maximizing their policy utility. While reporting of limitations, model re-use, evaluation practices, and stakeholder engagement improved over time, there is limited adoption overall. Overall, policy-claiming studies tended to demonstrate stronger evaluation practices and somewhat higher levels of stakeholder engagement, but model code is only publicly available in 38.70% of studies. These findings suggest that while ABMs proved valuable as exploratory tools, their potential as decision-support systems was only partially realized. Strengthening standards for transparency, evaluation, and interdisciplinary collaboration and stakeholder engagement remains essential for future pandemic preparedness.

**Acknowledgements**

**Supplements**

All supplementary material for this study can be found at [https://tinyurl.com/COVID-ABM-Review-Supplement](https://tinyurl.com/COVID-ABM-Review-Supplement).

Supplement 1: Search histories
Supplement 2: List of papers included in the study
Supplement 3: Data extraction form
Supplement 4: Additional sub-analysis of criteria
Supplement 5: PRISMA abstract checklist
Supplement 6: PRISMA article checklist
Supplement 7: Full dataset from extraction
Supplement 8: List of models and links

**References**

Anderson, W. (2021). The model crisis, or how to have critical promiscuity in the time of Covid-19. *Social Studies of Science*, *51*(2), 167–188. https://doi.org/10.1177/0306312721996053

Box, G. E. (1979). Robustness in the strategy of scientific model building. In *Robustness in statistics* (pp. 201–236). Elsevier. https://www.sciencedirect.com/science/article/pii/B9780124381506500182

Bui, T., & Lee, J. (1999). An agent-based framework for building decision support systems. *Decision Support Systems*, *25*(3), 225–237.

Castro, D. A., & Ford, A. (2021). 3D Agent-Based Model of Pedestrian Movements for Simulating COVID-19 Transmission in University Students. *ISPRS INTERNATIONAL JOURNAL OF GEO-INFORMATION*, *10*(8). https://doi.org/10.3390/ijgi10080509

de Andrade, B. S. S., Espindola, A. L. L., Faria Junior, A., & Penna, T. J. P. (2023). Agent-based model for COVID-19: The impact of social distancing and vaccination strategies. *INTERNATIONAL JOURNAL OF MODERN PHYSICS C*, *34*(10). https://doi.org/10.1142/S0129183123501322

DeAngelis, D. L., & Diaz, S. G. (2019). Decision-making in agent-based modeling: A current review and future prospectus. *Frontiers in Ecology and Evolution*, *6*, 237.

Edmonds, B., Le Page, C., Bithell, M., Chattoe-Brown, E., Grimm, V., Meyer, R., Montañola-Sales, C., Ormerod, P., Root, H., & Squazzoni, F. (2019). Different Modelling Purposes. *Journal of Artificial Societies and Social Simulation*, *22*(3), 6.

Eker, S. (2020). Validity and usefulness of COVID-19 models. *Humanities and Social Sciences Communications*, *7*(1), Article 1. https://doi.org/10.1057/s41599-020-00553-4

Ferguson, N. M., Donnelly, C. A., & Anderson, R. M. (2001). The Foot-and-Mouth Epidemic in Great Britain: Pattern of Spread and Impact of Interventions. *Science*, *292*(5519), 1155–1160. https://doi.org/10.1126/science.1061020

Fischer, L. S. (2016). CDC grand rounds: Modeling and public health decision-making. *MMWR. Morbidity and Mortality Weekly Report*, *65*. https://www.cdc.gov/mmwr/volumes/65/wr/mm6548a4.htm

Gilbert, N., Ahrweiler, P., Barbrook-Johnson, P., Narasimhan, K. P., & Wilkinson, H. (2018). Computational Modelling of Public Policy: Reflections on Practice. *Journal of Artificial Societies and Social Simulation*, *21*(1), 14. https://doi.org/10.18564/jasss.3669

Grefenstette, J. J., Brown, S. T., Rosenfeld, R., DePasse, J., Stone, N. T., Cooley, P. C., Wheaton, W. D., Fyshe, A., Galloway, D. D., Sriram, A., Guclu, H., Abraham, T., & Burke, D. S. (2013). FRED (A Framework for Reconstructing Epidemic Dynamics): An open-source software system for modeling infectious diseases and control strategies using census-based populations. *BMC Public Health*, *13*(1), 940. https://doi.org/10.1186/1471-2458-13-940

Grimm, V., Berger, U., DeAngelis, D. L., Polhill, J. G., Giske, J., & Railsback, S. F. (2010). The ODD protocol: A review and first update. *Ecological Modelling*, *221*(23), 2760–2768. https://doi.org/10.1016/j.ecolmodel.2010.08.019

Grimm, V., Railsback, S. F., Vincenot, C. E., Berger, U., Gallagher, C., DeAngelis, D. L., Edmonds, B., Ge, J., Giske, J., Groeneveld, J., Johnston, A. S. A., Milles, A., Nabe-Nielsen, J., Polhill, J. G., Radchuk, V., Rohwäder, M.-S., Stillman, R. A., Thiele, J. C., & Ayllón, D. (2020). The ODD Protocol for Describing Agent-Based and Other Simulation Models: A Second Update to Improve Clarity, Replication, and Structural Realism. *Journal of Artificial Societies and Social Simulation*, *23*(2), 7.

Holmdahl, I., & Buckee, C. (2020). Wrong but Useful—What Covid-19 Epidemiologic Models Can and Cannot Tell Us. *New England Journal of Medicine*, *383*(4), 303–305. https://doi.org/10.1056/NEJMp2016822

Hunter, E., & Kelleher, J. D. (2021). Adapting an Agent-Based Model of Infectious Disease Spread in an Irish County to COVID-19. *SYSTEMS*, *9*(2). https://doi.org/10.3390/systems9020041

Ioannidis, J. P. A., Cripps, S., & Tanner, M. A. (2022). Forecasting for COVID-19 has failed. *International Journal of Forecasting*, *38*(2), 423–438. https://doi.org/10.1016/j.ijforecast.2020.08.004
18

**List of Supplemental Materials:**